\documentclass{article}
\usepackage{spconf,amsmath,graphicx,cite,amsfonts,amssymb}
\usepackage[caption=false]{subfig}
\usepackage{booktabs} % For prettier tables

\title{UNSUPERVISED LEARNING OF DEEP FEATURES FOR MUSIC SEGMENTATION}
\name{Matthew C. McCallum}
\address{Gracenote Inc.}

\begin{document}

\ninept
\maketitle
\begin{abstract}
Music segmentation refers to the dual problem of identifying boundaries between, and labeling, distinct music segments, e.g., the chorus, verse, bridge etc. in popular music. The performance of a range of music segmentation algorithms has been shown to be dependent on the audio features chosen to represent the audio. Some approaches have proposed learning feature transformations from music segment annotation data, although, such data is time consuming or expensive to create and as such these approaches are likely limited by the size of their datasets. While annotated music segmentation data is a scarce resource, the amount of available music audio is much greater. In the neighboring field of semantic audio unsupervised deep learning has shown promise in improving the performance of solutions to the query-by-example and sound classification tasks. In this work, unsupervised training of deep feature embeddings using convolutional neural networks (CNNs) is explored for music segmentation. The proposed techniques exploit only the time proximity of audio features that is implicit in any audio timeline. Employing these embeddings in a classic music segmentation algorithm is shown not only to significantly improve the performance of this algorithm, but obtain state of the art performance in unsupervised music segmentation.
\end{abstract}
\begin{keywords}
Music information retrieval, Acoustic signal processing, Convolutional neural network, Deep learning
\end{keywords}

% Write sampling section
% Write results
% Write boundary detection equations
% Produce results figure / table

% Write conclusion
% Write abstract
% Produce network architecture figure
% Produce sampling distribution figure
% Produce empirical sampling positive examples figure
% Produce empirical sampling negative examples figure
% Produce SSM examples figure
% Clean references
% Clean equations / nomenclature
% Double check parameters specified in results (include CQT parameters?)
% Include algorithm fine-tuned for the SALAMI dataset?

% Add weighting by track length to positive sampling picture
% Add theoretical False positive rate to positive sampling picture marginalized over all segment lengths.
% Evaluate reference algorithms with all the same CQT configuration

\section{Introduction}
\label{sec:intro}

Music segmentation refers to the task of labeling distinct segments of music in a way that is similar to a human annotation. For example the chorus, verse, intro, outro and bridge in popular music.. The boundary between such segments may be due to a number of factors, for example, a change in melody or chord progression, a change in rhythm, changes in instrumentation, dynamics, key or tempo. This task is generally evaluated with two classes of metrics. The first class, boundary detection, refers to the ability of the algorithm to locate the locations of such boundaries in time. The second class, segment labelling, refers to the labelling of segments where two segments that are disconnected in time are labelled as same or different based on their perceptual similarity \cite{raffel2014, nieto2014, lukashevich2008}.

\subsection{Prior Work}
\label{subsec:priorwork}

A number of techniques exist in the literature that address either the boundary detection \cite{foote2000, mcfee2014OLDA, serra2014}, segment labelling \cite{nieto20142dfft} problems, or both simultaneously \cite{levy2008, mcfee2014, weiss2011, nieto2013}. Methods addressing solely the former metric typically involve the generation of a novelty function via a self similarity matrix (SSM) representation. Other approaches addressing both of the aforementioned metrics focus on clustering audio features based on characteristics that are expected to remain homogeneous within a given musical segment. Such clustering has been performed on the basis of timbral and harmonic features in the context of spectral clustering \cite{mcfee2014}, or in the context of time-translation invariant features such as HMM state histograms \cite{levy2008}. In addition, a number of feature transformations have been considered with respect to clustering approaches, including non-negative matrix factorizations (NMFs) of audio features \cite{nieto2013} or self-similarity matrices \cite{kaiser2010}, as well as learned features from labelled segmentation data \cite{mcfee2014OLDA}.

In the context of supervised deep learning some work has addressed the problem of music segmentation \cite{ullrich2014, schluter2014, grill2015}, where performance is likey bound by the limited annotated music segmentation data available. The largest effort to collect labeled music segmentation data is in the SALAMI dataset \cite{smith2011}, providing 2246 annotations of music structure in 1360 audio files. This amount of data is small in the context of deep learning, but promising results were obtained.

Recently, in the neighboring field of semantic audio, unsupervised approaches towards embedding audio features have attracted attention in the literature \cite{jansen2018, salamon2015, salamon2015feature}. In particular, the benefits of learning audio feature embeddings in an unsupervised manner have become apparent in \cite{jansen2018}. Because there is no requirement for labeled data in training such embeddings, they can be trained on much larger datasets and in turn, this has achieved impressive results when such pre-trained embeddings are employed in sound classification and query by example tasks \cite{jansen2018}. Despite the promising results obtained for tasks related to the identification of events in audio, little to no investigation has been made into the use of such unsupervised audio feature embeddings for music content specifically. 

\subsection{Contributions}
\label{subsec:contributions}

This work focuses on the unsupervised training of CNNs to obtain meaningful features for music segmentation methods. This is a natural progression of the application of modern machine learning methods to the problem of music segmentation for three reasons. 

Firstly, previous literature pays careful attention to obtaining features that are representative of musical characteristics that are perceptually important in identifying segment boundaries such as timbre or harmonic repetition \cite{foote2000, paulus2006, chai2006, marolt2006, jensen2007, weiss2011, pauwels2013, mcfee2014, serra2014, mcfee2014OLDA}. Whether or not a given feature is important in identifying a segment boundary or label may be genre or song specific and so a data based approach may be promising in either producing a representation that better generalizes across genres or in the least may be arbitrarily learned for each genre specifically. Several data based approaches have been investigated for the music segmentation problem \cite{mcfee2014OLDA, weiss2011, ullrich2014}, with little to no work in the context of unsupervised deep learning.

Secondly, musical content has a structure that may be exploited to further improve the machine learning methodologies employed in works such as \cite{jansen2018, salamon2015, salamon2015feature}. The characteristics that define music such as rhythm and harmonicity have not been exploited in these previous works. warranting some investigation in the context of music.

Finally, labeled data for the problem of music segmentation is notoriously time consuming and/or expensive to produce \cite{wang2017}, as such an unsupervised machine learning approach that can exploit large amounts of unlabeled music data is highly desirable. The approach in this paper investigates deep learning in an unsupervised framework, where only the time locality and a comparative analysis of time local data is exploited. Such an approach overcomes the necessity to hand annotate data, and hence may be scaled to the full extent of the available music data. No longer limited by the size of the training dataset, it is expected to generalize better than hand crafted features that rely on aspects such as timbre or repetition that may be specific to certain music genres.

\section{Audio feature embedding}
\label{sec:features}

The approach of here is to create an embedding for audio features by transforming an audio representation via a CNN into a domain that is representative of musical structure. These embeddings may be trained by employing a loss function, such as contrastive loss \cite{hadsell2006}, or triplet loss \cite{schroff2015}, that observes positive (similar) or negative (dissimilar) pairs or triplets of examples, where a triplet represents an anchor and both a positive example and a negative example. The triplet loss function is often shown to result in superior performance to contrastive loss, which has been argued to be due to the its relative nature \cite{wu2017}. That is, the triplet loss function is designed to provide a positive gradient with respect to increasing distance between positive pairs or the anchor and positive example of a triplet, relative to the distance between negative pairs, up to a given margin. As such, it is employed in this research.

If a transformation from an input feature, $x[q,k]$ (or equivalently $x$ for notational simplicity) to an embedding space, e.g., via a CNN, is described as $f : \mathbb{R}^{K\times Q} \rightarrow \mathbb{R}^D$, then a Euclidean triplet loss with a given margin, $\alpha$, and a mini-batch of training data consisting of the set $\mathcal{T}=\left\lbrace x^{c}_a, x^{c}_p, x^{c}_n \right\rbrace_{c=0}^{c=C-1}$ may be described as, 
\footnotesize
\begin{equation}
\mathcal{L}(\mathcal{T}) = \sum_{c=0}^{C-1}\left[ \left\lVert f\left( x^{c}_a \right) - f\left( x^{c}_p \right) \right\rVert^2_2 - \left\lVert f\left( x^{c}_a \right) - f\left( x^{c}_n \right) \right\rVert^2_2  + \alpha \right]_+ ,
\end{equation}
\normalsize
where subscript $a$, $p$ and $n$ represent an anchor, positive and negative input example respectively.

Ideally, in the context of music segmentation positive examples will be composed of audio feature representations from the same music segment as the anchor, while negative examples will represent audio features from distinct music segments. Thus, once trained, input features could be transformed via that CNN into a space where distinct clusters with respect to a Euclidean distance metric would represent distinct music segments. In an unsupervised context with no prior information about the music segmentation, exact selection of such examples is not possible, however strategies are described in Section~\ref{sec:sampling} that may improve this selection.

The properties that are significant in forming clusters representing music segments in the embedded space are learned from the data at the CNN input. These embedded features may be representative of any of the musical qualities mentioned in Section~\ref{sec:intro}, provided that these qualities are observable from the input features. Here, a constant-Q transform (CQT) \cite{schorkhuber2010} is used due to its ability to accurately represent transient and harmonic audio qualities, its translation invariant representation of harmonic structures, and the promising results observed for this feature specifically in \cite{nieto2016}.

A time window of CQT data providing $K$ frequency bins across $Q$ time windows forms a time frequency representation that may compose any of $x^{c}_a$, $x^{c}_p$ or $x^{c}_n$. In this work it is found significantly advantageous to synchronize the CQT analysis windows with the beat of the music. That is, if a beat at index $i$ occurs at time $b_i$, then $R$ CQT windows are analyzed centered at times $b_i + r(b_{i+1}-b_{i})/R$ for integers, $r\in\left\{0..R-1\right\}$. These beat synchronized CQT representations are then aggregated across $B$ beats providing $Q=BR$ time indices in the representation $x[q,k]$ at the input to the CNN. 

\section{Sampling}
\label{sec:sampling}

Motivated by the unreasonable effectiveness of  data in deep learning \cite{sun2017}, methods are proposed here that create noisy positive / negative examples exploiting the facts that a) musical segments form contiguous regions in a song's timeline, and b) each distinct musical segment label typically occurs for the minority of a song's timeline. Specifically, it is proposed to use the time proximity information implicit in a song's features - sampling features that occur close together or at a minimum distance apart for positive and negative examples respectively. That is, an anchor beat index, $i_a$, is selected via a uniform sampling of the beat indices in a given song, $\left\{0..L-1\right\}$. Thereafter, a positive beat index, $i_p$, is chosen uniformly sampled from beat indices $\left\{\max(i_a-\delta_p, 0)..\min(i_a+\delta_p, L-1)\right\}$, and a negative beat index $i_n$ is chosen as uniformly sampled across two regions, $\left\{\max(i_a-\delta_{n,max}, 0)..\max(i_a-\delta_{n,min}, 0)\right\}$, and $\left\{\min(i_a+\delta_{n,min}, L-1)..\min(i_a+\delta_{n,max}, L-1) \right\}$. An example of the positive and negative sampling distributions is shown in Fig.~1. Intuitively, this results in an embedding in which clusters represent features that frequently occur close together. This is useful for the problem of the structural segmentation of music as segment boundaries are typically infrequent enough to be described as rare events, at least in comparison to lengths of the segments themselves.

It is interesting to consider the rates of false positives and negatives that result from the aforementioned sampling paradigm. Upon selection of $i_a$ it may be denoted to fall somewhere in the $n$th musical segment of class $s$ and length $l_{s,n}$. If $\delta_p<l_{sep}$, where $l_{sep}$ is the minimum number of indices between $s$ and any identically labeled segment, then the probability of the positive example being selected from a distinct segment, i.e., a false positive, is, 

\begin{equation*}
P(\mathrm{FP}|l_{s,n} ; \delta_p) = \begin{cases} 
\frac{2\delta_p - l_{s,n}}{l_{s,n}} & l_{s,n} \leq \delta_p \\
\frac{\delta_p^2}{2l_{s,n}^2} - \frac{3\delta_p}{4l_{s,n}} + \frac{1}{2} & \delta_p < l_{s,n} < 2\delta_p \\
\frac{\delta_p}{4l_{s,n}} & 2\delta_p \leq l_{s,n} 
\end{cases}.
\end{equation*}
Similarly, if $\delta_{n,min}<l_{sep}$ and $\delta_{n,max} \geq L$, then the probabillity of a false negative is simply $\frac{\sum_m l_{s,m}}{L} - (1-P(\mathrm{FP}|l_{s,n} ; \delta_{n,min}))$.

In practice the aforementioned assumptions are realistic in many scenarios, but do not hold under all conditions. An empirical analysis of the rate of false positives is shown in Fig.~2. It is clear that with an increasing $\delta_p$, an increasing rate of false positives is observed, although it is important to note that smaller $\delta_p$ restricts the maximum observed time separation between the anchor and positive example. With musical phrases often lasting 16 beats, it is reasonable to set $\delta_p \geq 16$ to discourage distinct clustering of features within a single phrase. An empirical analysis false negatives is shown in Fig.~3. There it can be seen that for the datasets shown $\delta_{n,min} > 28$ and $\delta_{n,max} > 116$ result in relatively low false negative rates.

\begin{figure}[t]
\includegraphics[width=230pt]{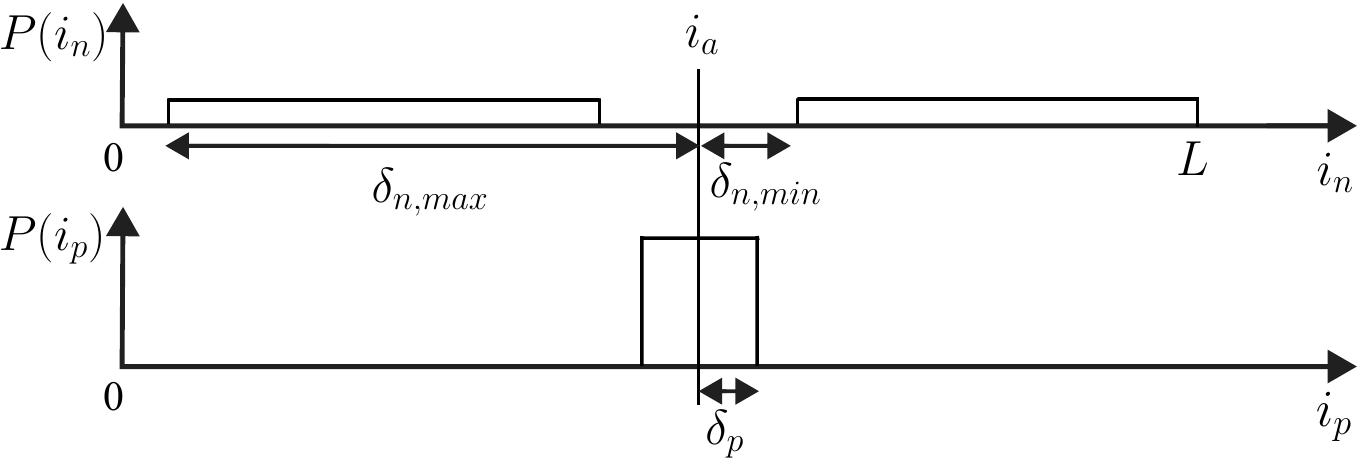}
\label{fig:sampling}
\vspace{-0.5em}
\caption{Sampling distributions for $i_n$ and $i_p$, for a given choice of $i_a$.}
\vspace{-1em}
\end{figure}

\begin{figure}[t]
\includegraphics[width=220pt]{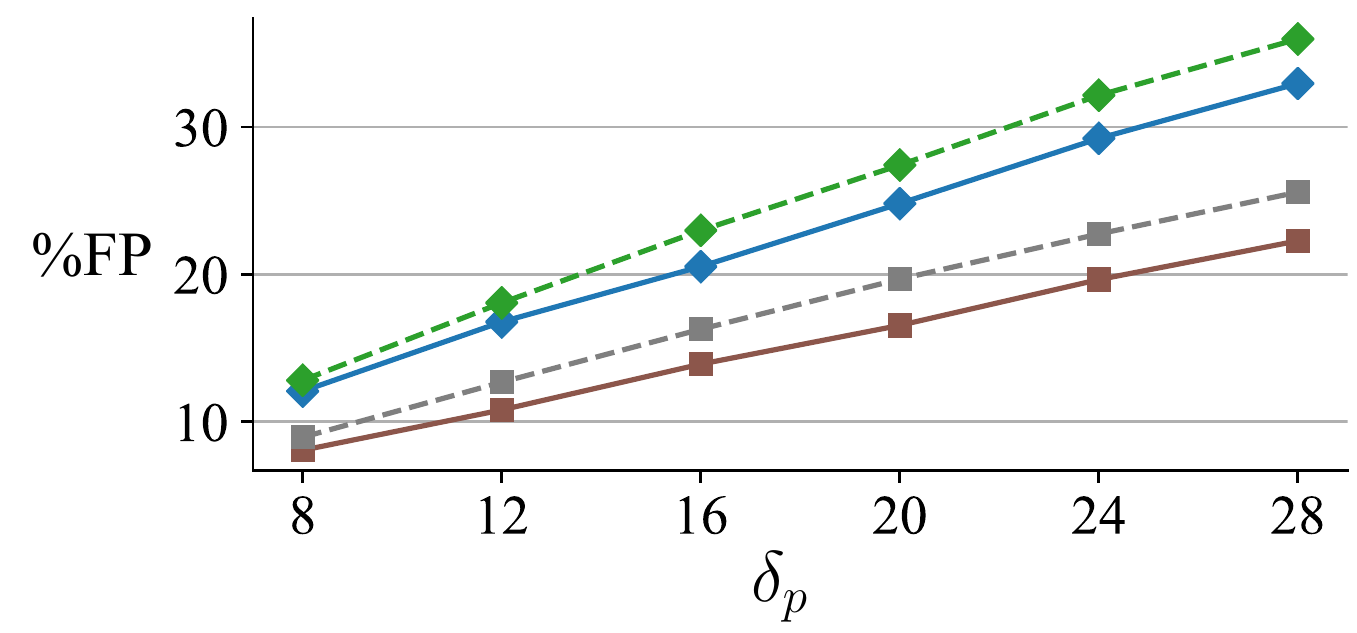}
\label{fig:FalsePositives}
\vspace{-0.5em}
\caption{Observed false positive (FP) rates in the BeatlesTUT ($\diamond$), and SALAMI ($\square$), datasets for varying $\delta_p$. Dashed lines show results with unbiased sampling, and solid lines employ the comparitive 2D FFT technique described here.}
\vspace{-1em}
\end{figure}

\begin{figure}[t]
\includegraphics[width=250pt]{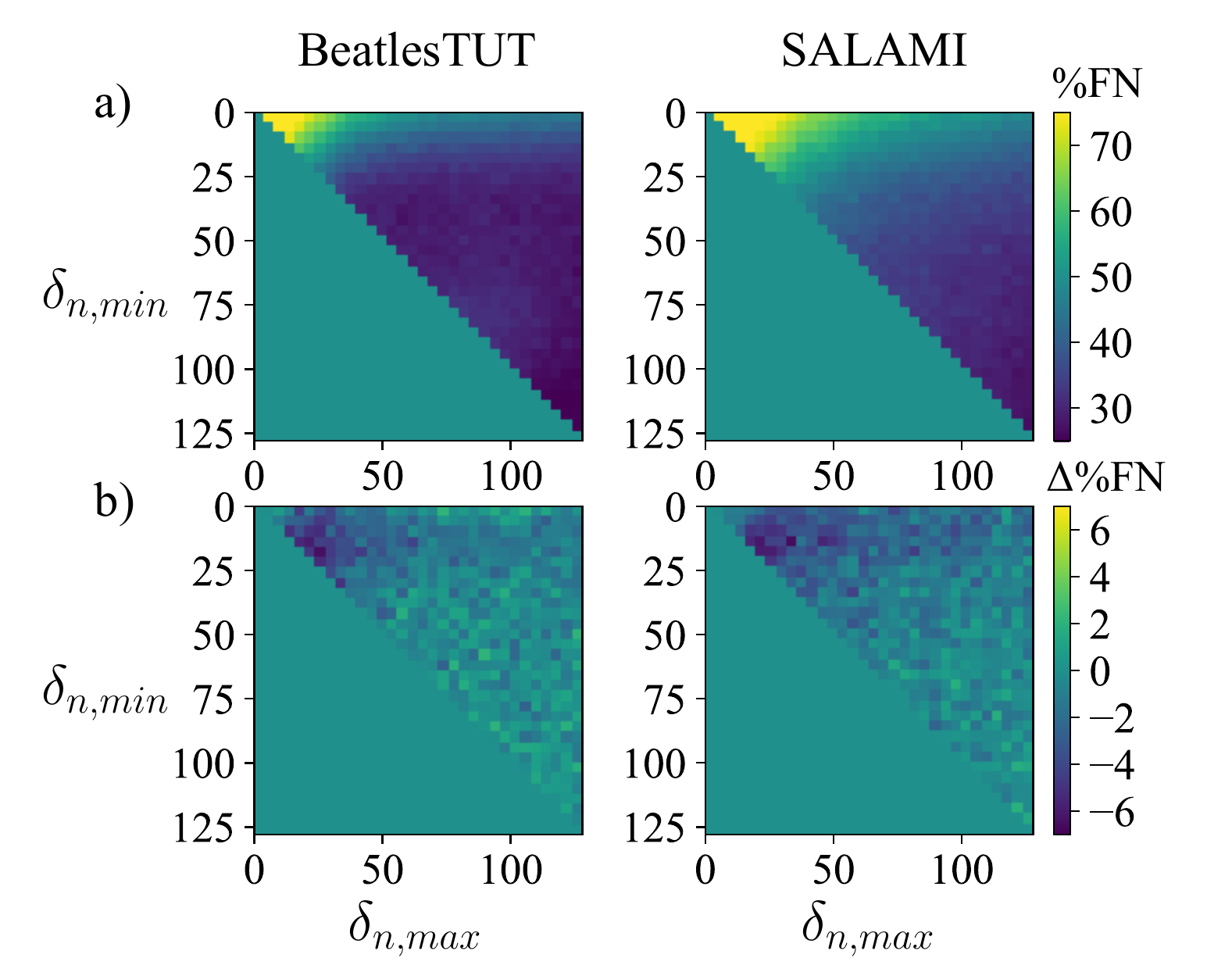}
\label{fig:FalseNegatives}
\vspace{-1.15em}
\caption{Observed false negative (FN) rates for various $\delta_{n,min}$ and $\delta_{n,max}$. Row (a) shows FN raters resulting from unbiased sampling, while row (b) represents the change in FN rates when employing the comparitive 2D FFT technique described in this paper.}
\vspace{-1em}
\end{figure}

While structural annotations are not available under the scope of unsupervised learning, it is interesting to ask whether analysis of signal features may be employed to decrease the false-negative or false-positive rate. It was shown in \cite{nieto20142dfft} that 2D Fourier magnitude coefficients of HPCPs can be a useful feature in segment labeling. It is proposed here to use a similar feature to inform the sampling of positive and negative examples. That is, for every selected $i_a$, a comparison of the log-amplitude of the 2D Fourier transform of the log-amplitude of 8 beat long CQT segments is considered. The Euclidean distances between these CQT segments centered at two times before $i_a$, i.e., $i_a-4$ and $i_a-16$, and two times after $i_a$, i.e., $i_a+4$ and $i_a+16$ are considered. The side (``before" where $i<i_a$ or ``after" where $i>i_a$) with minimum Euclidean distance is then assumed to be more likely in the same musical segment as $i_a$ and is chosen from which to sample $i_p$ while the opposite side may be chosen from which to sample $i_n$, in addition to the constraints, $\delta_p$, $\delta_{n,min}$ and $\delta_{n,max}$ above. The changes in false negative and false positive rates employing this sampling paradigm can be seen in Figures~2 and~3, respectively, where some improvement is observed.

\section{Boundary Detection}
\label{sec:boundaries}

To evaluate the effectiveness of the music embeddings described in this work, the problem of music segment boundary detection is considered. Perhaps the simplest and most well known method for music boundary detection is that of Foote \cite{foote2000}. While this has been surpassed in performance by several methods, e.g., \cite{serra2014, mcfee2014}, its simplicity makes it effective in demonstrating the utility of the audio embeddings proposed here, as will be seen in the results in Section~\ref{sec:results}.

The SSM of the proposed features is computed as,

\begin{equation}
S[i,j]=\left\lVert f\left( x_i[q,k] \right) - f\left( x_j[q,k] \right) \right\rVert^2_2
\end{equation}
where $x_i[q,k]$ and $x_j[q,k]$ correspond to beat synchronous CQT segments centered at beats $i$ and $j$, respectively. 

Note that this work endeavors to create features that are close with respect to Euclidean distance at any point within a music segment. If successful, the SSM of embedded features typically contains block structures as opposed to the path structures typically representative of repetition. These structures are evident in Fig.~4. It was found beneficial in practice to perform median filtering on $S[i,j]$ to produce $\bar{S}[i,j]$ which reduces noise in the distances between embedded features while maintaining the aforementioned structures.
% Could add a reference to explain the path structures mentioned above.

\begin{figure}[t]
\includegraphics[width=240pt]{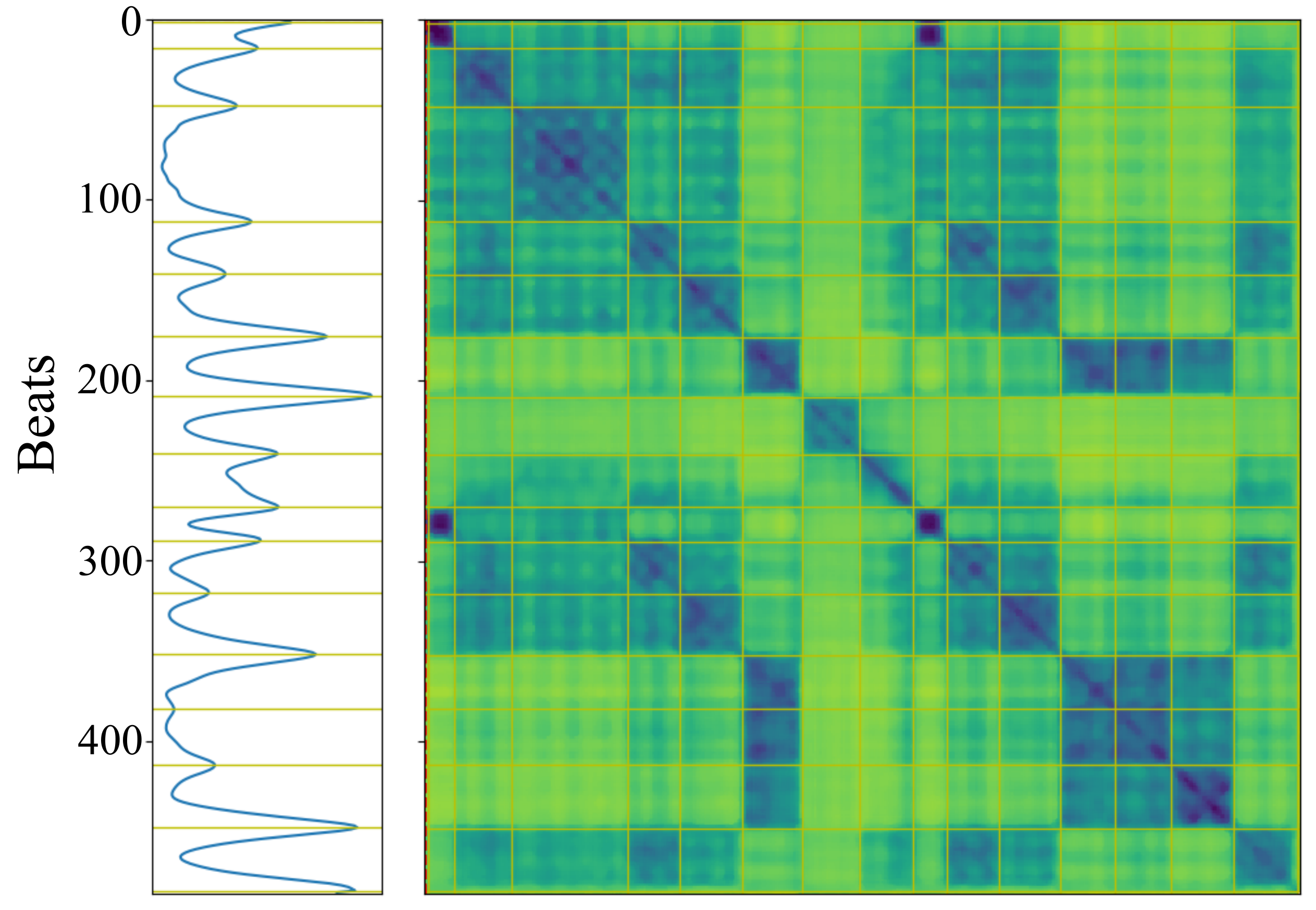}
\label{fig:SSM}
\vspace{-0.5em}
\caption{Example of median filtered SSM and novelty function for the song "Starlight" by Muse. On both, detected boundaries are marked with yellow lines.}
\vspace{-0.5em}
\end{figure}

To detect segment boundaries in $\bar{S}[i,j]$, a checkerboard kernel,

\begin{equation}
g[i,j]=\mathrm{sgn}(i) \mathrm{sgn}(j) e^{\frac{-(i-j)^2}{\sigma}}
\end{equation}
for $-\kappa \leq i,j \leq \kappa$ is convolved along the diagonal of the SSM,

\begin{equation}
\eta[\nu]=\sum_{i=-\kappa}^\kappa \sum_{j=-\kappa}^\kappa \bar{S}[\nu+i,\nu+j]g[\nu+i,\nu+j]
\end{equation}
producing the novelty function $\eta[\nu]$. Note that it was found advantageous to set $g[i,j]$ to $0$ where $\vert i \vert \leq 1$ or $\vert j \vert \leq 1$. Finally, boundaries are detected as peaks in the novelty function. Specifically, peaks at which the peak-to-mean ratio exceeds a given threshold, $\tau$, are selected as segment boundaries, i.e., where,
\begin{equation}
\frac{(2T+1)\eta[\nu]}{\sum_{t=-T}^T \eta[\nu+t]} > \tau.
\end{equation}

\section{Results}
\label{sec:results}

For evaluation, two datasets are considered: the BeatlesTUT dataset consisting of 174 hand annotated tracks from The Beatles catalogue \cite{paulus2006}, and the internet archive portion of the SALAMI dataset (SALAMI-IA) consisting of 375 hand annotated recordings \cite{smith2011}. The former dataset is perhaps the most widely evaluated in the music segmentation literature. The latter is employed here for two reasons, firstly the complete SALAMI dataset audio is not available to the author due to copyright restrictions, and secondly, the SALAMI-IA dataset is particularly interesting as it consists primarily of live recordings with many imperfections. It provides a dataset that is indicative of segmentation performance when there are mistakes either by musicians or recording engineers, resulting in imperfect repetitions and distorted or noisy audio in many cases.
% Can you improve the reference for the beatlesTUT dataset?

For comparison, two baseline algorithms are included in the results as specified and measured in \cite{nieto2016}, note that these algorithms too included beat-synchronized features. Firstly, the method of \cite{foote2000} is included as it most closely mirrors the algorithm of Section~\ref{sec:boundaries}. Secondly, the algorithm of \cite{serra2014} is widely evaluated as having the best performance with respect to unsupervised boundary detection in music segmentation. CQT features have been evaluated to provide superior performance in \cite{nieto2016}, and are used at the input to all algorithms, proposed and benchmark, in the results presented here. For the proposed algorithms, three methods are investigated, "Unsynchronized", employing a constant hop size of 3.6~ms between successive CQT windows; "Beat-Synchronized", employing 128 CQT windows centered at times linearly interpolated between successive beat markers (estimated during training and inference by an algorithm similar to \cite{klapuri2006}); and "Biased", employing the same features as the "Beat-Synchronized" approach, but using the 2D Fourier Transform comparison sampling described in Section~\ref{sec:sampling}. The CQT features use a minimum frequency of 40~Hz, 12 bins per octave and 6 octaves.
%FIGURE - NETWORK ARCHITECTURE - No normalization on input as we assume the same track all to be mixed at the same level. Include L2 normalization on output.

\begin{figure}[t]
\includegraphics[width=240pt]{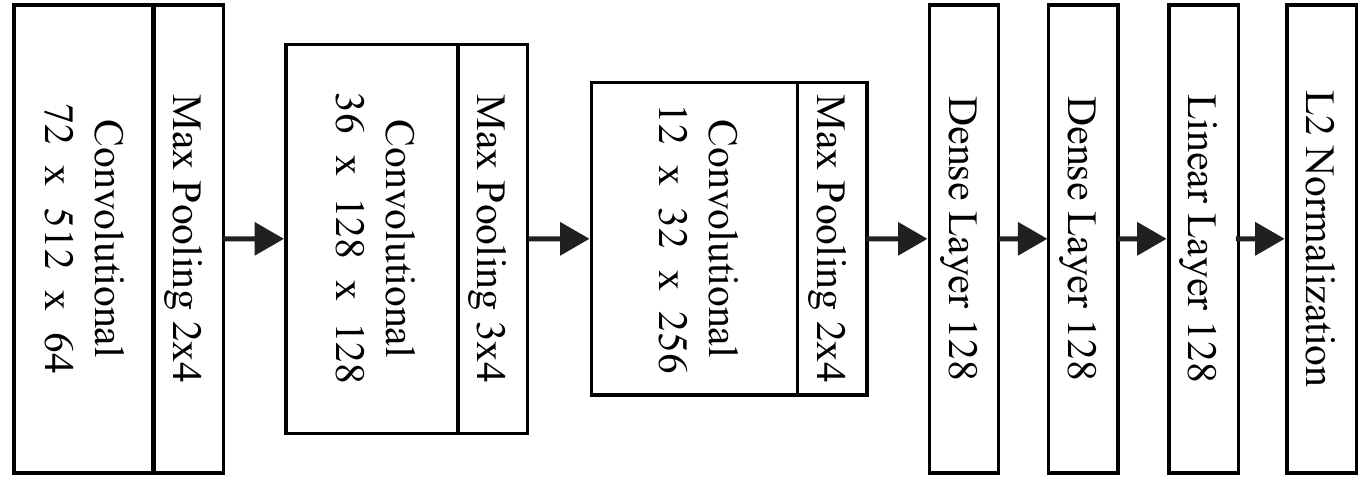}
\label{fig:CNN}
\vspace{-0.5em}
\caption{Architecture of the CNN employed in experiments. For each convolutional layer, dimensions represent time, frequency and channel respectively. Each convolutional and dense layer employ a ReLU activation. All convolutional layers employ $6\times4$ kernels.}
\vspace{-1em}
\end{figure}

The structure of CNNs employed in this paper is chosen to be simple enough so that it may be trained on most modern high performance GPUs. The architecture described in Fig.~5 was found to be effective and is employed in the experiments here. During training, mini-batches of size $C=96$ consisting of 16 triplets from each of 6 randomly selected tracks were formed in real-time by randomly selecting from 28,345 songs, excluding all songs in the SALAMI-IA and BeatlesTUT dataset. 256 min-batches form 1 epoch, and training took place over 240 epochs taking approximately 8 hours. The triplet margin was set to $\alpha=0.1$. Despite the observed error rates in Figures~\ref{fig:FalsePositives} and~\ref{fig:FalseNegatives}, \cite{schroff2015, wu2017} argue that it is the difficulty of separation between examples that is important, and so not all false positives and negatives are equal. In practice it was found that $\delta_p=16$, $\delta_{n,min}=1$ and $\delta_{n,max}=96$ provide optimal results.

For boundary detection, embedded features were observed at every beat for beat-synchronized approaches, or once every 0.2484 seconds for "Unsynchronized" (this is approximately twice per beat at a typical 120~BPM to ensure this method is not disadvantaged by any shortcoming in time resolution). SSM representations were median filtered using an 8x8 window. The checkerboard kernel was configured with $\sigma=18.5$ and $\kappa=40$ and for peak picking, the crest factor window size $T=10$ and threshold $\tau=1.35$ was employed.

For evaluation, the trimmed F-measure, precision and recall of the boundary detection hit rate at the 3 second tolerance level are employed \cite{raffel2014}. The evaluation of the proposed and reference algorithms for the BeatlesTUT and SALAMI-IA dataset are shown in Table~\ref{tab:BeatlesResults} and Table~\ref{tab:SALAMIResults}, respectively. It should be noted here that for the proposed algorithms, any selection of parameters was performed by observing results on the BeatlesTUT dataset only, and so the SALAMI-IA dataset displays the boundary detection algorithm's ability to generalize to unseen data.

It is interesting to see that simply by employing the proposed deep features in an algorithm similar to that of Foote \cite{foote2000}, such a method becomes competitive with the state of the art in unsupervised music segmentation. Furthermore, on the SALAMI-IA dataset, a significant performance improvement over the state of the art is observed without any additional parameter adjustment. This result might be postulated to be due to the poor quality of audio / music data in this portion of the SALAMI dataset. Because the algorithm of \cite{serra2014} is designed to detect changes in repetition patterns, when these patterns become imperfect, or corrupted by noise, a performance drop might be expected. In the proposed embedding, clustering of features is performed simply based on the time proximity of features observed from the training data, which contains many of the aforementioned imperfections providing some robustness.

\begin{table}[t]
  \vspace{-1em}
  \begin{center}
    \caption{Performance metrics for the BeatlesTUT dataset}
    \label{tab:BeatlesResults}
    \begin{tabular}{l|l|l|l}
      \toprule % <-- Toprule here
      \textbf{Algorithm} & \textbf{F-Measure} & \textbf{Precision} & \textbf{Recall}\\
      \midrule % <-- Midrule here
	  \cite{foote2000} & 0.503 $\pm$ 0.18 & 0.579 $\pm$ 0.21 & 0.461 $\pm$ 0.17 \\
      \cite{serra2014} & 0.651 $\pm$ 0.17 & 0.622 $\pm$ 0.19 & \textbf{0.708} $\pm$ 0.19 \\
      Unsynchronized & 0.597 $\pm$ 0.17 & 0.589 $\pm$ 0.19 & 0.625 $\pm$ 0.17 \\
      Beat-Synchronized & 0.648 $\pm$ 0.17 & 0.647 $\pm$ 0.20 & 0.677 $\pm$ 0.18 \\
      Biased Sampling & \textbf{0.662} $\pm$ 0.17 & \textbf{0.663} $\pm$ 0.20 & 0.691 $\pm$ 0.19 \\
      \bottomrule % <-- Bottomrule here
    \end{tabular}
  \end{center}
  \vspace{-2em}
\end{table}

\begin{table}[t]
  \begin{center}
    \caption{Performance metrics for the SALAMI-A dataset}
    \label{tab:SALAMIResults}
    \begin{tabular}{l|l|l|l}
      \toprule % <-- Toprule here
      \textbf{Algorithm} & \textbf{F-Measure} & \textbf{Precision} & \textbf{Recall}\\
      \midrule % <-- Midrule here
	  \cite{foote2000} & 0.446 $\pm$ 0.17 & 0.457 $\pm$ 0.21 & 0.483 $\pm$ 0.19 \\
      \cite{serra2014} & 0.493 $\pm$ 0.17 & 0.454 $\pm$ 0.20 & 0.595 $\pm$ 0.19 \\
      Unsynchronized & 0.497 $\pm$ 0.16 & 0.429 $\pm$ 0.18 & 0.653 $\pm$ 0.15 \\
      Beat-Synchronized & \textbf{0.535} $\pm$ 0.15 &  \textbf{0.491} $\pm$ 0.20 &  \textbf{0.660} $\pm$ 0.16 \\
      Biased Sampling & 0.533 $\pm$ 0.16 & \textbf{0.491} $\pm$ 0.21 & 0.656 $\pm$ 0.16 \\
      \bottomrule % <-- Bottomrule here
    \end{tabular}
  \end{center}
  \vspace{-2em}
\end{table}

\section{Conclusion}
\label{sec:conclusion}

In this work, methods for the unsupervised training of music embeddings were investigated with respect to their utility in the task of music segmentation. In particular, it was shown that by employing such embeddings in a traditional music segmentation algorithm, the performance of this algorithm can obtain state of the art performance. It was found that a common musical feature, rhythm, may be exploited in beat-synchronized sampling (and in the 2D Fourier Transform comparitive sampling of Section~\ref{sec:sampling}) to further improve performance.

\vfill\pagebreak

\bibliographystyle{IEEEbib}
\bibliography{refs}

\end{document}